\documentclass[fleqn,12pt]{wlscirep}
\usepackage[utf8]{inputenc}
\usepackage[T1]{fontenc}

\usepackage{booktabs}
\usepackage{multirow}
\usepackage{longtable}
\usepackage{xr-hyper}
\usepackage[charter,cal]{mathdesign}
\usepackage{bm}
\makeatletter
\newcommand*{\addFileDependency}[1]{
  \typeout{(#1)}
  \@addtofilelist{#1}
  \IfFileExists{#1}{}{\typeout{No file #1.}}
}
\makeatother

\newcommand*{\myexternaldocument}[1]{%
    \externaldocument{#1}%
    \addFileDependency{#1.tex}%
    \addFileDependency{#1.aux}%
}
\myexternaldocument{SI}

\newcommand{\dl}{{\mathscr{L}}}
\newcommand{\xs}{{\mathbf{x}}}
\newcommand{\fs}{{\mathbf{f}}}
\newcommand{\Fs}{{\mathbf{F}}}
\newcommand{\thetas}{{\bm{\theta}}}
\newcommand{\dxs}{{\dot{\mathbf{x}}}}
\newcommand{\dx}{{\dot{x}}}
\newcommand{\dy}{{\dot{y}}}
\newcommand{\dD}{{\dot{D}}}
\newcommand*\diff{\mathop{}\!\mathrm{d}}

\title{
Integral Bayesian symbolic regression for optimal discovery of governing equations from scarce and noisy data\\
}

\author[a]{Oriol Cabanas-Tirapu}
\author[a]{Sergio Cobo-Lopez}
\author[b]{Savannah E. Sanchez}
\author[c]{Forest L. Rohwer}
\author[a,*]{Marta Sales-Pardo}
\author[a,e,*]{Roger Guimerà}

\affil[a]{Department of Chemical Engineering, Universitat Rovira i Virgili, 43007 Tarragona, Catalonia}
\affil[b]{Department of Microbiology and Immunology, Virginia Commonwealth University School of Medicine. 1101 East Marshall Street P.O. Box 980678, Richmond, VA 23298-0678, USA}
\affil[c]{Department of Biology, San Diego State University, San Diego, CA, USA}
\affil[e]{ICREA, 08007 Barcelona, Catalonia}

\affil[*]{Corresponding authors: Marta Sales-Pardo (E-mail: marta.sales@urv.cat); Roger Guimerà (E-mail: roger.guimera@urv.cat)}

\begin{abstract}
Understanding how systems evolve over time often requires discovering the differential equations that govern their behavior. Automatically learning these equations from experimental data is challenging when the data are noisy or limited, and existing approaches struggle, in particular, with the estimation of unobserved derivatives. Here, we introduce an integral Bayesian symbolic regression method that learns governing equations directly from raw time-series data, without requiring manual assumptions or error-prone derivative estimation. By sampling the space of symbolic differential equations and evaluating them via numerical integration, our method robustly identifies governing equations even from noisy or scarce data. We show that this approach accurately recovers ground-truth models in synthetic benchmarks, and that it makes quasi-optimal predictions of system dynamics for all noise regimes. Applying this method to bacterial growth experiments across multiple species and substrates, we discover novel growth equations that outperform classical models in accurately capturing all phases of microbial proliferation, including lag, exponential, and saturation. Unlike standard approaches, our method reveals subtle shifts in growth dynamics, such as double ramp-ups or non-canonical transitions, offering a deeper, data-driven understanding of microbial physiology. 
\end{abstract}
\begin{document}

\flushbottom
\maketitle



\section*{Introduction}

Closed-form mathematical models are crucial to understand the behavior of natural and human-made systems across scientific and engineering disciplines. They provide explicit analytical descriptions of the mechanisms that govern phenomena and enable prediction of future outcomes, control, and optimization. Closed-form models are particularly valuable because they are interpretable, computationally efficient, and facilitate the derivation of fundamental principles. Symbolic regression (also known as equation discovery) plays a key role in elucidating such governing equations by leveraging data-driven machine learning approaches to identify mathematical expressions that best describe observed relationships\cite{cava2021,makke2024}. Unlike traditional regression methods, symbolic regression explores the space of possible mathematical forms, combining machine learning with symbolic mathematics, to uncover parsimonious and physically meaningful equations. This makes it a powerful tool for uncovering hidden patterns and deriving interpretable models from empirical data, bridging the gap between statistical and mechanistic modeling \cite{evansmachinescience,wang2023,cornelio23}.

Within the symbolic regression framework, identifying the governing equations of dynamical systems in the form of differential equations \cite{brunton16,quade16} presents especial challenges, particularly when it comes to dealing with the derivatives. These derivatives are usually not directly observable and must be estimated numerically from the observed data. However, numerical estimation of derivatives introduces biases and leads to spurious correlations, which significantly impact the performance of most symbolic regression techniques when applied to learning differential equations \cite{pynumdiff2020}.

Perhaps the best known and most widely used method for learning governing equations of dynamical systems from data is sparse identification of nonlinear dynamical systems (SINDy) \cite{brunton16,desilva2020}. SINDy uses sparsity-promoting techniques to identify parsimonious differential equations in the form of linear combinations of predefined sets of library functions. However,  SINDy is very sensitive to noise because of its reliance on numerically estimated derivatives. To alleviate this shortcoming, several SINDy extensions have been proposed, including those based on smoothing of derivatives \cite{he2022} and estimating confidence intervals for model parameters \cite{egan24}, those based on Bayesian estimates of the parameters \cite{niven19,niven24}, those based on bagging and ensembles of models to increase robustness \cite{fasel2022,Mangan2017}, and, most significantly for the purpose of our work, those based on weak \cite{reinbold2020,messenger2021} and integral formulations \cite{schaeffer2017} of the equation discovery problem. In the latter, integrated versions of the differential governing equation are used to completely avoid derivative estimation. Furthermore, to mitigate the issues raised by the need of expert input to predefine library functions, combinations with machine learning methods have also been proposed to address the problem \cite{cranmer20,liu23,hu25}.

Here, we introduce an integral Bayesian approach to identify governing equations from scarce and noisy data. Like integral versions of SINDy and other approaches that use numerical integration of the differential equations \cite{stolle07,mangiarotti12,omejc24}, it does not require numerical estimation of derivatives. However, unlike SINDy-based approaches, our approach does not require that the differential equation is a linear combination of previously known library functions; rather, it can have any arbitrary form. Additionally, also unlike SINDy, it does not require any hyperparameter tuning. Finally, because it is based on rigorous probabilistic arguments, our approach provides optimal treatment of the uncertainty associated with data scarcity and observational noise.

To validate our approach, we test it in well-known model dynamical systems, in a regime (scarce data and high noise) in which even state-of-the-art approaches typically fail to identify the correct governing equation. We find that the integral Bayesian approach does discover the true generating models in this regime, up to high levels of noise and even with only tens of data points. We also discuss the transition by which, due to observational noise, governing equations become unlearnable by any method \cite{fajardo-fontiveros23}; and show that the integral Bayesian approach makes optimal predictions of system trajectories both above and below this transition. Finally, we apply our approach to learn the governing equations of bacterial growth, and show that the models discovered perform significantly better than logistic and Gompertz models typically used to model these phenomena.

\section*{Results}
\subsection*{Probabilistic formulation of the problem of discovering governing equations from noisy data}

Let us consider a system whose evolution $\xs(t)$ is governed by the differential equation
\begin{equation}
\dxs = \fs^*(\xs, \thetas)\,, \quad \xs \in \mathbb{R}^n \;, \qquad \fs^*: \mathbb{R}^n \to \mathbb{R}^n\;, 
\label{eq:ode}
\end{equation}
where $\thetas$ are some fixed but unknown parameters. If the time derivatives $\dxs$ are observed directly, along with the coordinates $\xs$ themselves, the problem of discovering the governing (vector) function $\fs^*(\xs, \thetas)$ can be addressed by standard symbolic regression. In this scenario, one typically assumes that the observations of the derivatives have some measurement error
\begin{equation}
\dx_{ik} = f^*_i(\xs_k, \thetas) + \epsilon_{ik} \,, 
\end{equation}
where $\dx_{ik}$ is the $k$-th observation of the time derivative of the $i$-th component of $\xs$, $f^*_i$ is the $i$-th component of $\fs$, $\xs_k$ is the $k$-th observation of $\xs$, and $\epsilon_{ik}$ is a Gaussian random variable with zero mean and unknown variance $\sigma$. Under these conditions, the complete and optimal \cite{fajardo-fontiveros23} probabilistic solution of the problem is encapsulated in the posterior distribution $p(f_i|\dD_i)$ \cite{guimera20,fajardo-fontiveros23}, which gives the probability that an arbitrary function $f_i(\xs, \thetas)$ is the true governing function $f^*_i(\xs, \thetas)$, given the observed data $\dD_i = \{(\dx_{ik}, \xs_k)\}$. (Note that the dot on $\dD_i$ just indicates that time derivatives are observed directly.) Without loss of generality, this posterior can be written as \cite{guimera20,fajardo-fontiveros23}
\begin{equation}
p(f_i | \dD_i) = \frac{1}{Z} \exp{\left[-\dl\left(f_i, \dD_i\right)\right]} \,,
\label{eq:dl}
\end{equation}
where $\dl(f_i, \dD_i)$ is the description length of model $f_i$, measured in nats, and results from integrating the likelihood of $f_i(\xs, \thetas)$ over its parameters $\thetas$ (Methods). Under mild assumptions \cite{schwarz78,guimera20}, it can be approximated as
\begin{equation}
\dl(f_i, \dD_i) = \frac{B(f_i, \dD_i)}{2} - \log p(f_i) \,,
\label{eq:bic}
\end{equation}
where $B(f_i, \dD_i)$ is the Bayesian information criterion \cite{schwarz78} of model $f_i$, and $p(f_i)$ is a suitable prior distribution over models \cite{guimera20}. The normalization constant in Eq.~\eqref{eq:dl} (or partition function) is $Z = \sum_{f_i} \exp{\left[-\dl\right(f_i, \dD_i)]}$, where the sum is over all possible functions $f_i(\xs, \thetas)$. We call  Bayesian machine scientist (BMS)  any algorithm that selects models by maximizing the posterior in Eq.~\eqref{eq:dl} (or, equivalently, minimizing the description length in Eq.~\eqref{eq:bic})\cite{guimera20}. When the derivatives $\dxs$ are observed directly, the BMS is Bayesian-optimal for learning the corresponding governing equations from data \cite{fajardo-fontiveros23}.

However, in most situations of interest, only the coordinates $\xs$ are observed and measured, but not the time derivatives $\dxs$. The simplest approach in this situation is to estimate the derivatives $\dxs$ numerically from $\xs$ using finite differences, and then proceed as in a regular symbolic regression problem, that is, as if the derivatives themselves had been observed. However, this approach leads to problems arising from the instability and spurious correlations that result from numerical estimates of the derivatives. Smoothing the derivatives reduces the instabilities, but introduces further spurious correlations, which, as we show below, are also problematic.

To solve this limitation, we present an integral formulation of the probabilistic BMS framework outlined so far. We start by integrating Eq.~\eqref{eq:ode} between the initial time $t=0$ and some arbitrary time $t$
\begin{equation}
\xs(t) = \xs_0 + \int_{0}^{t} \fs^*(\xs, \thetas) \, \diff t' \equiv \Fs^*(t, \thetas, \xs_0) \,.
\end{equation}
For any given component $x_i$, we can then assume that at time $t_k$ the data are generated as
\begin{equation}
x_{ik} = F_i^*(t_k, \thetas, \xs_0) + \epsilon_{ik} \,,
\end{equation}
and the posterior over integral governing functions $F_i$ (and, thus, over the corresponding differential governing functions $f_i$) is
\begin{equation}
    p(f_i | D) = p(F_i | D) = \frac{1}{Z} \exp{\left[-\dl(F_i, D)\right]} \,,
    \label{eq:idl}
\end{equation}
where the data are now a set of observations of the times and coordinates $D=\{(t_k, \xs_k)\}$, but not the time derivatives. The description length is as in Eq.~\eqref{eq:bic}, with the prior calculated on the differential function $f_i$ but the Bayesian information criterion calculated by comparing $F_i$ to the integrated data $x_{ik}$. Therefore, in this formulation there is no need to estimate the derivatives at any point. The integral governing function $F_i$ corresponding to a given governing function $f_i$ may or may not have a closed-form expression, but it can always be estimated numerically \cite{stolle07,mangiarotti12,omejc24} for given parameter values $\thetas$ and initial conditions $\xs_0$ (which can be regarded as additional parameters of the integral governing function), so that these parameters can be estimated and the description length $\dl(F_i, D)$ can be computed (Methods).
\begin{figure*}[t!]
\centering
\includegraphics[width=0.75\textwidth]{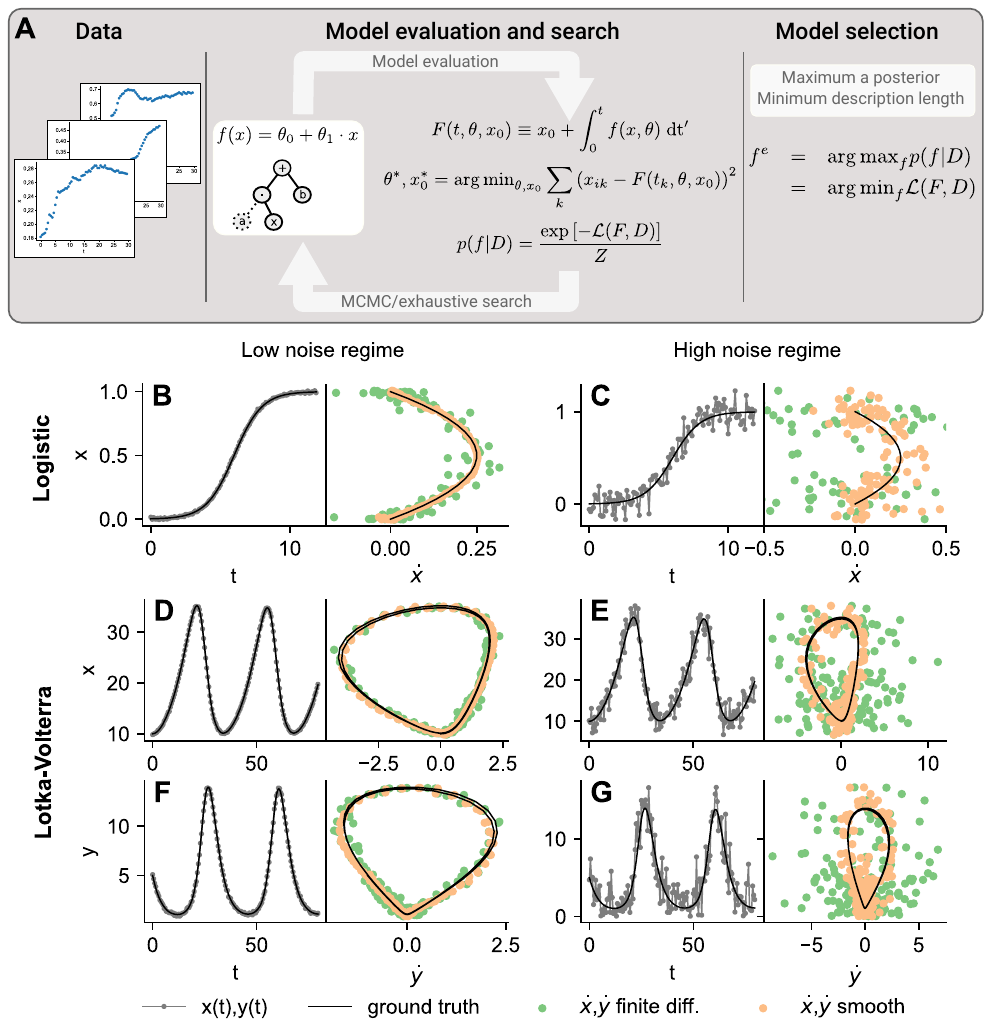} 
\caption[B]{\textbf{Integral Bayesian symbolic regression and benchmark data.} \textbf{(A)} Schematic representation of the approach. We start with scarce and noisy measured data $D$ from a system driven by the equation $\dx = f^*(x, \thetas)$. Given the observed data and any model $f$, we can evaluate the posterior probability $p(f|D)$ of the model without needing to estimate numerically the derivatives $\dx$. This involves optimizing model parameters $\thetas$ and initial conditions $x_0$ on the integrated form, and calculating the description length $\dl$ (see text). We select the model with the highest posterior (minimum description length), either by searching exhaustively within a predefined set of models or by sampling models through MCMC. \textbf{(B-G)} Left panels show the noisy synthetic data used in our validations, which we represent with gray lines; the black line corresponds to the noiseless ground truth behavior. Right panels show the phase space (measured variable against its derivative), with green dots representing the finite difference estimations of the derivative, yellow dots representing the smoothed estimate, and black lines representing the ground truth. \textbf{(B-C)} Logistic model. \textbf{(D-G)} Lotka-Volterra model. \textbf{(B, D, F)} Low noise regime. \textbf{(C, E, G)} High-noise regime.
}
\label{fig:fig0}
\end{figure*}
%
In this formulation, which we call integral Bayesian machine scientist (I-BMS), the most plausible governing function $f_i$ is the maximum a posterior or, equivalently, the one that minimizes the description length $\dl(F_i, D)$ of the integral governing function.

\subsection*{The integral Bayesian machine scientist recovers ground-truth governing equations up to high levels of noise}

To evaluate the ability of the integral probabilistic approach to identify ground-truth governing equations, we start by generating synthetic data with varying levels of noise and number of observations, and for two different systems: the one-dimensional logistic equation and the two-dimensional Lotka-Volterra system (Fig.~\ref{fig:fig0}; Methods).
In each scenario, we select the model $\fs^e$ whose integral governing function $\Fs^e$ minimizes the integral description length $\dl(\Fs, D)$ as defined in Eq.~\eqref{eq:idl}, and compare this model to the governing equation that truly generated the data. Each comparison is performed over 40 datasets to determine the frequency with which the approach converges to the exact ground-truth expression.

We benchmark these results against two sets of algorithms. First, we compare with the (non-integral) BMS  given by Eq.~\eqref{eq:dl} under two conditions: using finite-difference estimates of the derivatives $\dxs$ (FD-BMS), and smoothing those estimates by means of the derivative of a polynomial fit to the time series \cite{pynumdiff2020} (SD-BMS).
Second, we compare to ensemble-SINDy\cite{brunton16,fasel2022} in its weak/integral formulation \cite{reinbold2020,schaeffer2017,messenger2021,desilva2020}. SINDy uses sparse regression techniques to identify the relevant terms in a prescribed linear combination of certain library functions. In the context of discovering governing equations, these library functions are typically powers or products of powers of the components $x_i$ (for example, $x_1$, $x_1^3$ or $x_1 x_2^2$). Ensembling (ESINDy) adds robustness to SINDy by considering ensembles of models (obtained by resampling the data) as opposed to one single model \cite{fasel2022}.

Weak and integral formulations of SINDy and ESINDy are similar in spirit to the I-BMS, in the sense that they also integrate the SINDy models of the governing equations to avoid numerical estimation of the derivatives; but, like SINDy and ESINDy, they are restricted to consider models that are linear combinations of certain predefined functions. Therefore, for the comparisons in this section, we similarly restrict our approach to consider the same linear combinations allowed by weak ESINDy (W-ESINDy). However, instead of using sparse regression to choose among models (that is, to choose which terms belong to the model), we exhaustively evaluate the description lengths $\dl(F_i, D)$ of each allowed model\cite{bartlett2024}, and select the model $\fs^e$ whose corresponding $\Fs^e$ has components such that $F_i^e = \arg \min_{F_i} \dl(F_i, D)$, that is, that minimize the integral description length.


\begin{figure*}[t!]
\centering
\includegraphics[width=0.9\textwidth]{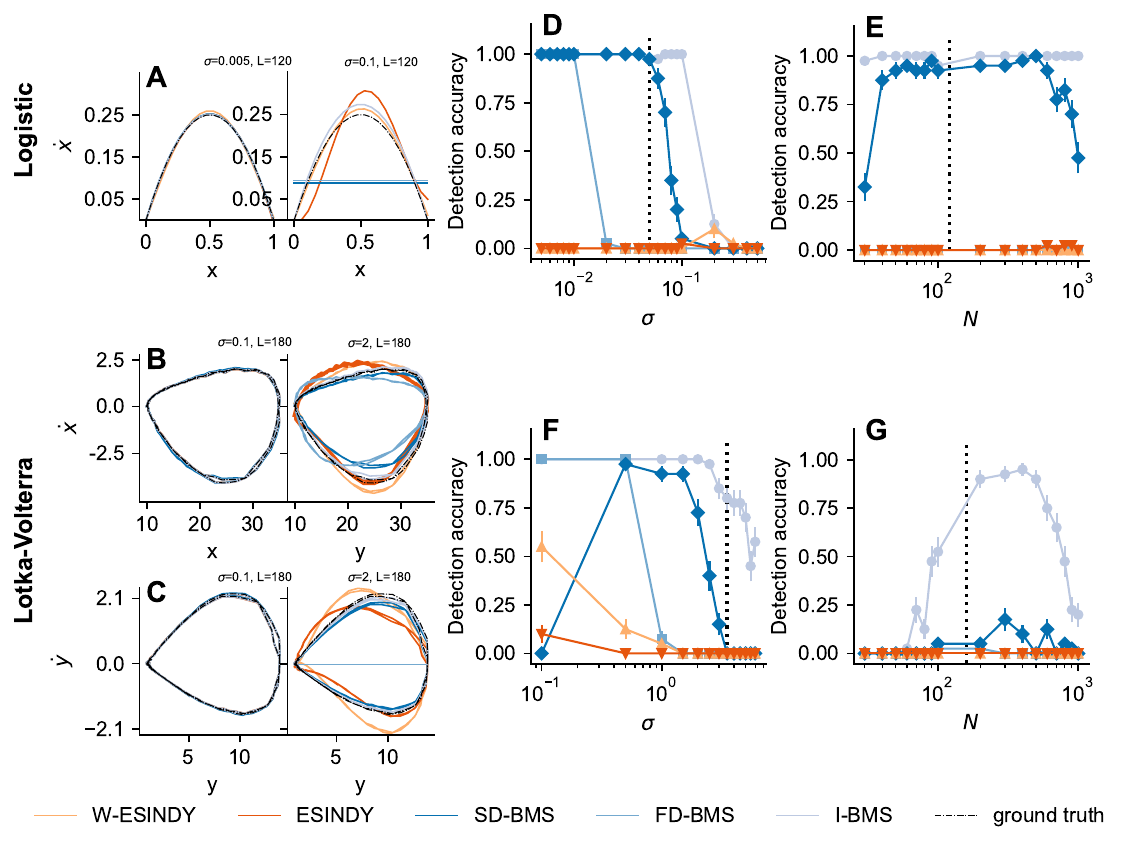}
\caption[B]{\textbf{Validation on synthetic data.} We generate synthetic data using the logistic and Lotka-Volterra models, with different levels of noise and different number of data points.
We then explore exhaustively the space of all polynomial expressions (see ``Exhaustive search of linear terms'' in Methods) for $f_i(\xs, \thetas)$, and use the BMS to select the model with the shortest description length. We do this for the integral BMS (I-BMS), as well as for the standard BMS with finite difference-estimated derivatives (FD-BMS) and with smoothed derivatives (SD-BMS); and we benchmark these algorithms against ensemble SINDy (ESINDy) and weak ensemble SINDY (W-ESINDy), with the same library functions as the BMS.
\textbf{(A-C)} Phase space trajectories predicted by the governing equations obtained using each of the approaches for one particular realization of the noise. Left panels correspond to the low-noise regime, while right panels correspond to the high-noise regime.
\textbf{(D-G)} To quantify the ability of each approach to identify the true governing equation, we show the detection accuracy, that is, the fraction of times that the true governing equation is exactly recovered (each data point corresponds to an average over 40 datasets $D$):
\textbf{(D)} as a function of noise level, for the logistic model and fixed number of observed points ($N = 120$); \textbf{(E)} as a function of the number of points, for the logistic model, fixed noise ($\sigma = 0.05$) and fixed total time range; \textbf{(F)} as a function of noise level, for the Lotka-Volterra model and fixed number of observed points ($N = 180$); \textbf{(G)} as a function of the number of points, for the Lotka-Volterra model, fixed noise ($\sigma = 3.5$) and fixed time spacing between consecutive observations.
}
\label{fig:detection_logistic}
\end{figure*}

For the range of noise and data sparsity levels considered, we find that W-ESINDy and ESINDy approximate the ground-truth governing equation relatively well (Fig.~\ref{fig:detection_logistic}A-C)), but are rarely able to identify it exactly when the function library is expanded beyond quadratic terms (Fig.~\ref{fig:detection_logistic}D-G; Supporting Text). 
%
Bayesian symbolic regression methods identify the ground-truth governing equations to different degrees in this situation. The regular BMS (Eq.~\eqref{eq:dl}) with finite-differences derivatives (FD-BMS) yields the poorest performance, being the most sensitive to noise (Fig.~\ref{fig:detection_logistic}D,F) and to data scarcity (Fig.~\ref{fig:detection_logistic}E,G). Smoothing of the derivatives (SD-BMS) leads to a much more robust identification, but sometimes generates serious problems for low levels of noise (Fig.~\ref{fig:detection_logistic}F) or abundant data (Fig.~\ref{fig:detection_logistic}E). In these situations, the probabilistic approach is overconfident about the data quality and tends to overfit. Altogether, the integral Bayesian symbolic regression (I-BMS) leads to the best results, that is, maximum resilience to noise and minimum requirements on the number of data points.

\subsection*{Fundamental limits of governing equation discovery and optimality of the integral approach}

We have shown how the BMS and especially the I-BMS clearly outperform state-of-the-art approaches in the identification of governing equations from noisy data. To make the comparison meaningful, this benchmarking was limited to the space of library functions considered by SINDy and its variants (in our case, polynomials). We now go a step further and investigate: (i) to what extent is it possible to identify governing equations when we lift the restriction of considering only linear combinations of certain library functions; (ii) how close are the identified governing equations to the ground truth, in terms of their dynamics.  

To answer these questions, we run a BMS that uses Markov chain Monte Carlo \cite{guimera20,reichardt20,fajardo-fontiveros23} (MCMC) to sample governing equations from the posterior distributions given by Eq.~\eqref{eq:dl}, in the non-integral formulation, and by Eq.~\eqref{eq:idl}, in the integral formulation. Note that the sampling process is not restricted to any specific form of the governing function $\fs(\xs, \thetas)$. For a given dataset $D$, the best estimate of the governing function is the $\fs^e(\xs, \thetas)$ with the shortest description length sampled by the BMS (integral or non-integral, depending on the case). We then define the ground truth as learnable for a given dataset $D$ if the description length of $\fs^e(\xs, \thetas)$ is larger or equal than that of the ground truth governing function  $\fs^*(\xs, \thetas)$. Conversely, a model is unlearnable if $\fs^e(\xs, \thetas)$ has a shorter description length than the ground truth governing function for that dataset. In that case $\fs^e(\xs, \thetas)$ is more plausible than the true model because $p(\fs^e|D) > p(\fs^*|D)$ and the ground-truth function cannot possibly be identified as the true model from the data alone \cite{fajardo-fontiveros23}.                                                                                                                                                                                                                                                                                                                                                                                                                                                                                                                                                                                                                                                                                                                                                                                                                                                                                                                                                                                                                                                                                                                                                                                                                                   


\begin{figure*}[t!]
\centering
\includegraphics[width=0.7\textwidth]{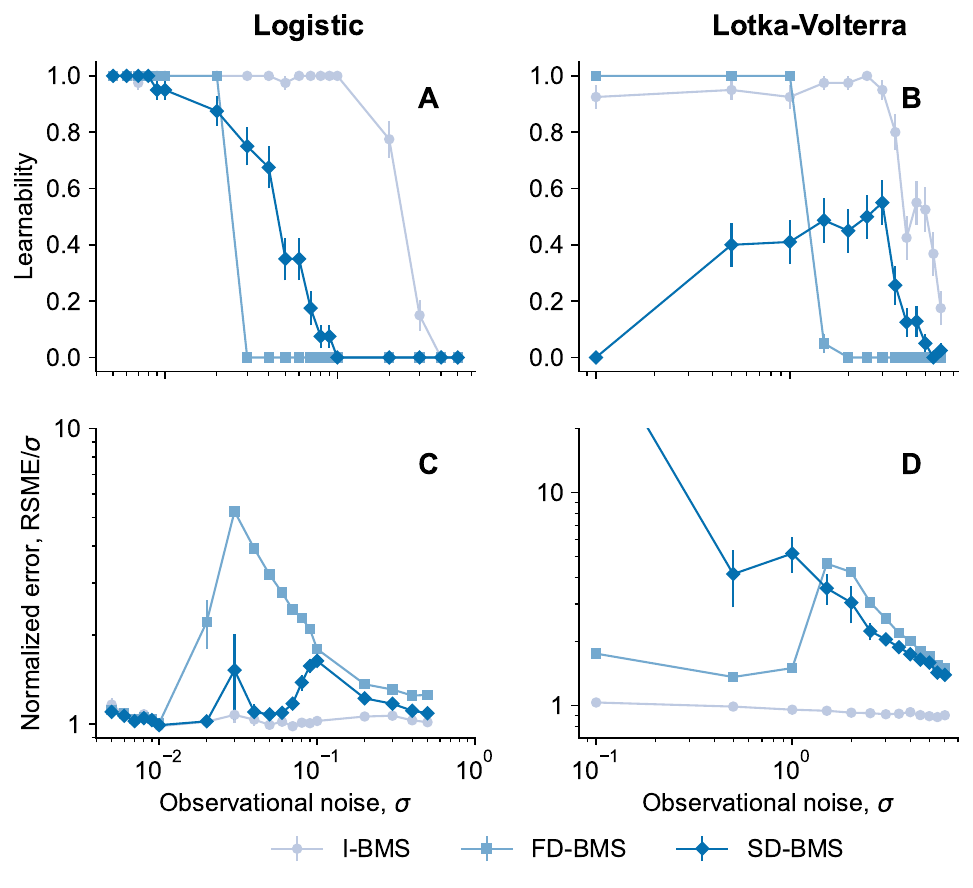} 
\caption[B]{\textbf{Learnability and model predictive accuracy across noise levels.}
We generate synthetic data for the logistic and Lotka-Volterra models, as in Fig.~\ref{fig:detection_logistic}. We then use MCMC to sample models from the posterior $p(f_i|D)$ (for each dataset, we run two independent MCMC processes with 3,000 steps each, except for the I-BMS on Lotka-Volterra data, for which we use 4,000 steps), and consider the most plausible model (equivalently, the model with the minimum description length). All points are averages over 40 datasets $D$. \textbf{(A-B)} Learnability as a function of noise level for the logistic and Lotka-Volterra datasets, respectively.
\textbf{(C-D)} Root mean squared error (RMSE) between the ground truth data $x(t)$ and the predictions of the minimum description length model $x^e(t)$, normalized by the noise level $\sigma$ for the logistic and Lotka-Volterra datasets, respectively.}
\label{fig:learnabilty}
\end{figure*}
As we show in Fig.~\ref{fig:learnabilty}A-B, the learnability of non-integral formulations is sensitive to noise, and the ground-truth governing function becomes unlearnable at low levels of noise. As observed before, estimating derivatives with smoothing leads to more resilience to noise than working directly with finite differences, but that comes at the cost of considerably reducing learnability at lower levels of noise. This, again, is due to the fact that smoothing makes data look more reliable (less fluctuating) than it really is, which leads the probabilistic approach to be overconfident and, thus, to overfit. As in the previous section, the integral formulation outperforms the non-integral approaches, and makes the ground truth learnable up to very high levels of noise.

To assess the predictive performance of all methods (regardless of whether the ground-truth function is learnable or not), we calculate the root mean squared error (RMSE) of the predictions $\xs^e(t)$ of the most plausible model $\fs^e(\xs, \thetas)$ with respect to the observed values of $\xs(t)$, where the predictions are given by integrating the governing function

\[
\xs^e(t) = \xs^e_0 + \int_{0}^{t} \fs^e(\xs, \thetas) \, \diff t'\,.
\]
Because $\xs(t)$ has a measurement error $\sigma$, optimal predictions are such that ${\rm RMSE}/\sigma = 1$. In Fig.~\ref{fig:learnabilty}C-D, we show that the models identified by the I-BMS are close to this optimal behavior for all levels of noise. Above the learnability threshold, this does not mean that the true model has been identified, but rather that any error related to the identification of incorrect models is small compared to the measurement error. In any case, our results suggest that, at least for the two systems considered, the I-BMS finds governing equations that are optimally predictive. By contrast, non-integral formulations lead to suboptimal predictions, especially in the transition region between the learnable and the unlearnable phases. 

\subsection*{Discovering governing equations for bacterial growth}

Having validated the predictive ability of the I-BMS on synthetic data, we apply it to a dataset comprising empirical growth curves of bacteria in various media, so as to obtain an equation governing bacterial growth.
%
%
We use six data sets containing data from laboratory experiments on different species\cite{cuevas2016,jove52854}: \textit{C. sedlakii}, \textit{E. coli}, \textit{E. aerogenes}, \textit{K. pneumaniae}, \textit{P. vulgaris} and \textit{S. aureus}. The bacteria were grown for 30 to 31 hours, and optical density OD$_{600}$ measurements of their populations were taken every 10 to 15 minutes. Each species was cultivated on 96 different substrates; here we only consider substrates that yielded substantial growth (exponential growth followed by a saturation phase). In total, we kept 29, 40, 24, 35, 28, and 13 substrates, respectively for each of the species. 
The training set comprises half of the experiments of five of the six bacterial species, randomly selected, whereas the test set includes the other half of the experiments on the five bacterial species and all the experiments on the remaining species.

We use the integral BMS to identify the most plausible model using MCMC, exploring model space with different inductive biases. First, we explore model space without imposing any constraints on the form of the mathematical model. Second, given the understanding that bacteria replicate through division, we consider a physics-informed I-BMS that focuses on models that incorporate a linear growth term, $\dot{B} = rB + g(B)$, where $B$ is the bacterial population size, $r$ is the growth rate and $g(B)$ is the arbitrary function that we aim to identify. Finally, we consider a scenario where growth rate is itself a function of bacterial population $h(B)$, so we ask the I-BMS to explore models including a term $\dot{B}= \alpha \,B\, h(B)$, where $\alpha$ is an arbitrary scaling parameter (see Methods). We run three independent sampling processes, each with 3,000 MCMC steps, and select the governing function with the minimum description length for each run.

We compare the results of the I-BMS to two classic and widely used models for bacterial growth: the logistic growth mode~\cite{wachenheim03} and the Gompertz model~\cite{wang24}. 
The logistic growth model is defined as
\begin{equation*}
    \frac{dB}{dt}=rB \bigg( 1-\frac{B}{K} \bigg) ~ ,
\end{equation*}
where, $B$ is the population size (in practice, the optical density gives a measure of $B$ shifted by a quantity, which can be interpreted as the optical density measured when there are no bacteria in the plate). The first term $r\,B$ gives rise to the exponential growth of bacterial population with a growth rate $r$, the maximum growth rate when nutrients are unlimited. The factor $1-B/K$ introduces the competition among bacteria for nutrients, so that the larger the population the larger the competition, and $K$ is the carrying capacity, that is, the theoretical maximum concentration of bacteria achieved in the steady state. 

The Gompertz model
\begin{equation*}
\frac{dB}{dt} = rB \log\left(\frac{K}{B}\right) = rB \left( \log K  -\log B \right) \,,
\end{equation*}
can be interpreted in a similar fashion, 
the same exponential growth as the logistic model and a logarithmic term that reduces growth when the population approaches the carrying capacity $K$. 

Recent studies that attempt to model the growth of cell populations in microbial communities use a generalized logistic model \cite{richards59} that interpolates between the logistic and Gompertz curves, making it possible for sublinear growths observed in microbial communities \cite{camacho-mateu25}. Nonetheless, we decided to compare against the classical models typically used for population growth of single cell types.


\begin{figure*}[t!]
\centering
\includegraphics[width=.9\textwidth]{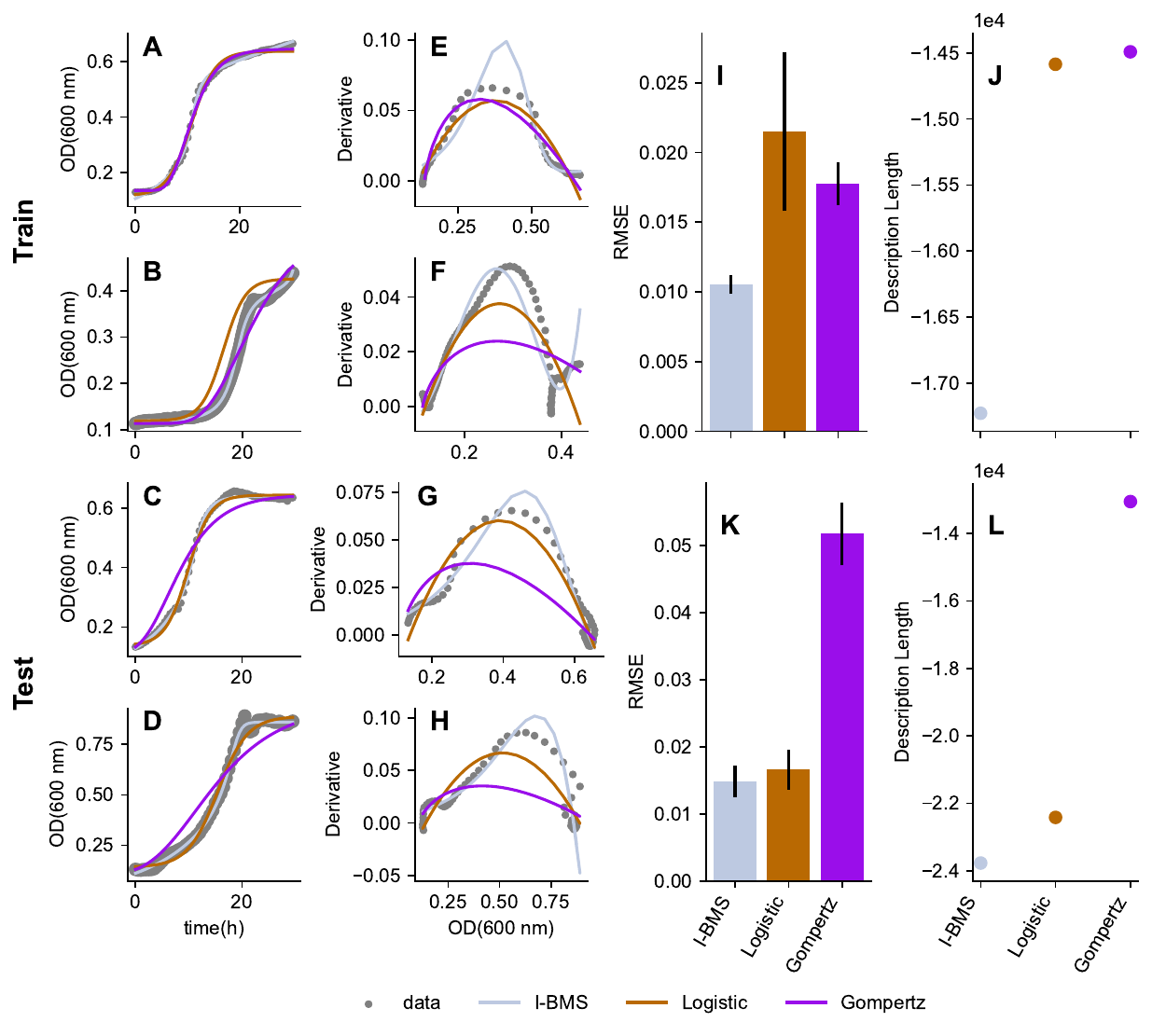} 
\caption[B]{\textbf{I-BMS model and reference growth models for bacterial growth.} We show results for two bacteria-substrate pairs from the training set, and two from the test set. \textbf{(A-D)} Empirical groth curves and numerically integrated curves $x^{e}(t)$ for each model. \textbf{(E-H)} Derivative values plotted against different measured optical densities for each model.  \textbf{(I,K)} Root mean squared error (RMSE) of the integrated curve relative to the observed data, computed for all datasets in the training and test sets, respectively. \textbf{(J-L)} Description length of the models for all training and test datasets.
}
\label{fig:bacterial_growth}
\end{figure*}
We find that the most plausible model identified by the I-BMS has the form
\begin{equation}
    \frac{dB}{dt} = B\,r\, \left(1 + c_{0}\left(c_{1} B  e^{c_2 B}\right)^{B^{3}}\right) \,,
\end{equation}
where $r$, $c_0$ $c_1$ and $c_2$ are model constants, that are different for each bacterial species and growth medium in the training set (that is, we find a single model for all datasets, but allow for model parameters to be dataset-specific\cite{reichardt20}). Formally, like in the logistic and Gompertz models, this model has two terms with opposite effects, as we find that $c_0$ and $r$ have opposite signs. The first term is proportional to growth, and is typically associated to exponential growth for $r>0$. The second term is a competition term typically reflecting competition for nutrients and other effects that limit growth. However, the dependence of this term on population size is more complicated than on benchmark models. Still, similarly to the other models, there is limiting population size for which $\frac{\diff B}{\diff t}=0$, although this carrying capacity cannot be expressed as a simple combination of parameters. Importantly, this higher complexity of the competition term increases model expressiveness, as this model outperforms the logistic and Gompertz models at describing bacterial growth (Fig.~\ref{fig:bacterial_growth}). Indeed, the I-BMS model successfully captures the dynamics of bacterial growth in all phases (lag phase, growth phase, and stationary phase), even allowing for non-standard behaviors in the lag and growth phases (double ramps, faster and slower growths, etc.) (Fig.~\ref{fig:bacterial_growth}A-D). Besides fitting the empirical data accurately, the model does not overfit spurious fluctuations. By contrast, the benchmark models often struggle to correctly capture the initial phase and the exponential growth phase (Supplementary Figs.~S1 and S3).
Furthermore, the I-BMS model provides a reasonable estimation of the empirical derivative curve (Fig.~\ref{fig:bacterial_growth}E-H), despite the fact that the empirical derivative is not used in any way during training and model selection. The benchmark models often fail, again, to approximate the derivative curve (Supplementary Figs.~S2 and S4).

In Figs.~\ref{fig:bacterial_growth}I-L, we show the quality metrics for both the training and test datasets. The I-BMS model demonstrates superior performance in both root mean squared error (although the difference is not significant when compared to the logistic model on test data) and description length, indicating a more parsimonious representation of the observed data.

\section*{Discussion}

Learning the differential equations governing the behavior of dynamical systems is fundamental, not only to predict their evolution, but also to elucidate the relevant underlying mechanisms. In recent years, this problem has been increasingly addressed through symbolic regression, which aims to automatically infer mathematical expressions that describe a system's dynamics. However, existing methods have struggled with overcoming two main limitations: fitting numerically-estimated derivatives---which introduces bias and amplifies noise; and restricting search space of  to a narrow predefined set of model structures---which simplifies the search but risks missing the relevant models. Recent deep learning-based approaches alleviate these limitations but have common caveats in that they need to define heuristic cost functions and heuristic search algorithms, and it is rarely tested under which circumstances they attain the desired (ground-truth) results.

Here, we have introduced an integral formulation of existing Bayesian symbolic regression approaches. The integral formulation addresses the first limitation mentioned above by numerically integrating candidate models rather than numerically estimating derivatives. By working directly with observed or measured system primitives, we avoid the amplification of noise typically introduced by numerical differentiation, which allows for more accurate model evaluation and increases the chances of recovering the true underlying dynamics. Consequently, the models discovered are more often correct and better aligned with the system's real behavior. Indeed, we find that, for synthetic data, the integral Bayesian approach recovers the true governing equations in regions where, due to noise or lack of data, all benchmark approaches fail.

The integral formulation also addresses the second limitation in that it is able to explore the whole space of symbolic expressions, without being restricted to linear combinations of predefined library functions. Therefore, our framework flexibly accommodates complex closed-form equations and is able to propose intricate models that are inaccessible to methods such as SINDy or to other classical methods for system discovery, which rely on linear combinations of library functions. We have confirmed this by studying empirical data of bacterial growth. For this system we discover a relatively simple but nonlinear differential equation, which is a more parsimonious description of the empirical data than existing models. 

Last but not least, our integral Bayesian approach is built on solid probabilistic arguments about data and model uncertainty and is thus Bayes optimal. This means that, if models where drawn from a known prior distribution $p(f)$, then our approach would lead to optimal selection of models and optimal predictions of behavior. In practice, we do not know the prior distribution. However, our experiments for synthetic data for which we have ground truth dynamics show that our approach makes optimal predictions about system dynamics, that is, predictions that have as little error as one may reasonably expect. Therefore, we argue that the proposed approach marks a significant step forward in data-driven discovery of governing equations and demonstrates the potential of integral Bayesian methods in advancing the automation of scientific modeling.

\section*{Methods}

\subsection*{Synthetic data}
We generated synthetic datasets for the one-dimensional logistic equation and the two-dimensional Lotka-Volterra model.
For the logistic model we use the integrated equation:
\begin{equation}
    x(t) = \frac{A}{B + e^{C - D t}} \,
\label{Logisic_eq}
\end{equation}
where $A=1$ , $B=1$, $C=0$ and $D=1$. To generate data according to the model,  we evaluated Eq.~\eqref{Logisic_eq} for $t \in [-6,6] $ with a step $\delta t = 0.1$, so that the number of observed points is $N=120$. For experiments with different noise levels, we generated 40 data sets for each Gaussian noise level $\mathscr{N}(0,\sigma)$, where $\sigma$ is the standard deviation of the distribution.
In experiments with a constant noise level but varying number of points $N$, we evaluated the logistic equation for $t \in [-6, 6]$ with datasets comprising equally spaced points, so that the larger the number of points the smaller $\delta t$. For each value of $N$, we generated 40 datasets, as before.

The Lotka-Volterra model is defined as a system of coupled differential equations
\begin{equation}
\begin{cases}
    \dx = ax-bxy\\
    \dy= dxy - cy
\end{cases}
\end{equation}
with $a=0.1$ , $b=0.02$, $c=0.02$ and $d=0.4$. We integrate the equations from $t_0=0$ to $t_1=80$ with a time step of $\delta t=0.1$ and initial condition $x_0=10$, $y_0=5$. Then, we generated 40 datasets for the different levels of noise by adding Gaussian noise $\mathscr{N}(0,\sigma)$.

\paragraph{Differentiation methods}
For numerical differentiation we use the central difference method. This involves computing the derivative at each observed time point $t_i$ using the values at the preceding and following time points, so that
\begin{equation}
x'(t_i) = \frac{x(t_{i+1}) - x(t_{i-1})}{2h} \; ,
\end{equation}
where $x'(t)$ is the numerical derivative, $x(t)$ is the value of the observable at time $t$, and $h=t(i)-t(i+1)$ is the time step. For the first and last data points, we use the forward and backward finite difference methods, respectively.

To obtain smoothed derivatives, we use the Python package \textit{PyNumDiff}\cite{pynumdiff2020}, which involves fitting a polynomial to the data and differentiating the fitted curve. Specifically, we used a second-order polynomial and a window size of 21 data points to smooth the data before computing the derivative.

\subsection*{MCMC Model Sampling}

To sample the posterior distribution over the space of models, we employ a Markov Chain Monte Carlo (MCMC) approach. The objective is to explore the space of possible mathematical models $\fs$, and identify the model $\fs^e$ with the highest posterior probability, which corresponds to the one that minimizes the description length.  

The general procedure follows the standard BMS approach \cite{guimera20}. At each MCMC step, we attempt an expression update followed by a parallel tempering swap process. We begin by setting an initial expression as the starting configuration, typically a one-node tree. This initial node is usually a parameter but may also be a variable. It is important to note that the operations that we can use in the expression tree are: exp(), pow2(), pow3(), -, +, *, / and **.
To ensure a better sampling of model space, we use parallel tempering. To that end, we clone the initial model (at temperature $T_0=1$) and assign a temperature  $T_k=1.02^k$ for $k \in [1,20]$ to each replica. 
 At each MCMC step, we propose a formula modification for each replica. This involves randomly altering the tree structure and computing the corresponding posterior of the modified expression. The description length associated with the posterior depends on whether we use the standard BMS \cite{guimera20} approach in Eq.~\eqref{eq:dl} or the integrated I-BMS formulation Eq.~\eqref{eq:idl}.  

Following the expression update, we propose a series of swaps adjacent replicas at $(T_k,T_{k+1})$. Finally, at the end of each MCMC step, we compare the description length of the model at $T_0$ and update the minimum description length model if a lower value is found.  

\paragraph{Initial Parameter Guess}  
In the I-BMS approach, we first fit the numerical derivative to obtain an initial estimate for the parameters of the differential equation. This estimate is then used as an initial guess for the optimization of the numerically integrated equations.  

\paragraph{2-Dimensional I-BMS}  
In the case of the 2D I-BMS, our objective is to identify a system of differential equations. Since the differential equations for each variable are coupled, their estimation must be handled simultaneously, both during parameter fitting and when computing the sum of squared errors.  
In each MCMC step, we first update the states for one variable across all temperature levels before proceeding to the other variable. When updating a single variable and computing the description length, we first determine the optimal parameters. These parameters are adjusted such that the numerical integration of the system of equations minimizes the discrepancy with the observed data.  
Next, we compute the sum of squared errors on the trajectory of the updated variable. If the proposed modification to $\fs$ is accepted, we update the optimal parameters for both equations in the system.


Note that at a fixed fr replica $k$ with $\fs_k=(f_x,f_y)$, we propose changes for $f_x$ and $f_y$ separately, but that the computation of the description length associated to the proposed model implies the calculation of a new set of parameters for both  $f_x$ and $f_y$. For swaps between adjacent replicas in the temperature sequence we also propose separate swaps for $f_x$ and $f_y$.


\subsection*{Exhaustive search of linear terms}

We explored polynomial expressions up to a given degree and up to a given number of terms. For each dataset, we evaluated all possible linear combinations using both numerical differentiation and integration approaches. Sepcifically, our search space was:

\begin{itemize}
\item \textbf{Logistic Equation:} Linear terms up to order 4 ($S = {1, x, x^2, x^3, x^4}$), generating polynomials with combinations up to 4 terms.
\item \textbf{Lotka-Volterra Model:} Linear terms up to order 3 ($S = {1, x, y, x^2, y^2, xy, x^2y, xy^2}$), generating polynomials with combinations up to 4 terms.
\end{itemize}

\subsection*{Physics-informed models}

For modeling bacterial growth, we use the I-BMS approach to explore arbitrary models that incorporate mathematical constraints derived from phenomenological principles. For instance, we could enforce that component $i$  includes an additive linear term so that $f_i=a\,x_i + f_{1i}({\bf x})$. In implementation terms, each BMS instance maintains an expression tree $f_{1i}$ that does not explicitly include these constraints but the full model needs to be used to compute the description length. 
Note that in order to assess whether the models found using these constraints have a lower description length than models obtained without constraints, we must use the full model $f_i$ (and not $f_{1i}$) to compute the prior probability $p(f_i)$.




\bibliography{ode}

\section*{Acknowledgments}
This research was supported by project PID2022-142600NB-I00 (MS-P and RG), and FPI grant PRE2020-095552 (OC-T) from MCIN/AEI/10.13039/501100011033, and by the Government of Catalonia (2021SGR-633) (MS-P and RG).

\section*{Author contributions statement}

OCT, MSP and RG designed the research. OCT wrote code and performed experiments. OCT, SCL, MSP and RG analyzed all results. SCL, SES and FLR collected and processed data and analyzed results on bacterial growth. All authors wrote the paper.

\section*{Data and code availability}
The datasets and code to replicate the experiments are available from \href{https://github.com/ocabanas/Integral_BMS_Governing_Equations}{Github}.

\section*{Competing interests}
The authors declare no conflict of interest.

\end{document}